\documentstyle[aps,twocolumn,epsf,floats]{revtex}
\begin{document}
\wideabs{
\title{LATERAL TUNNELING THROUGH THE CONTROLLED BARRIER BETWEEN EDGE
CHANNELS IN A TWO-DIMENSIONAL ELECTRON SYSTEM}
\author{ A.A.~Shashkin, V.T.~Dolgopolov, E.V.~Deviatov,
B.~Irmer  \dag, A.G.C.~Haubrich \dag,
J.P.~Kotthaus \dag,
M.~Bichler  \ddag, W.~Wegscheider \ddag }
\address{Institute of Solid State Physics, Chernogolovka, Moscow
District 142432, Russia \protect \\
\protect \dag Ludwig-Maximilians-Universit\"at,
  Geschwister-Scholl-Platz
1, D-80539 M\"unchen, Germany  \protect \\
\protect \ddag Walter Schottky Institut, Technische Universit\"at
M\"unchen, D-85748 Garching, Germany  }
\date{}
\maketitle

\begin{abstract}
We study the lateral tunneling through the gate-voltage-controlled
barrier, which arises as a result of partial elimination of the donor
layer of a heterostructure along a fine strip using an atomic force
microscope, between edge channels at the depletion-induced edges of a
gated two-dimensional electron system. For a sufficiently high
barrier a typical current-voltage characteristic is found to be
strongly asymmetric and include, apart from a positive tunneling
branch, the negative branch that corresponds to the current
overflowing the barrier. We establish the barrier height depends
linearly on both gate voltage and magnetic field and we describe the
data in terms of electron tunneling between the outermost edge
channels.
\end{abstract}
\pacs{72.20 My, 73.40 Kp}
}

Recently there has arisen much interest in the lateral tunneling into
the edge of a two-dimensional electron system (2DES), which is
related not only to the problem of integer and fractional edge states
in the 2DES but also to that of resonant tunneling and
Coulomb-blockade \cite{pal,is,man,vk,chang,bev,wharam}. The tunneling
regime was identified by exponential dependences of the measured
current on either source-drain voltage \cite{pal,is,man,vk} or
magnetic field \cite{chang}. For producing a tunnel barrier a number
of methods were used: (i) gate voltage depletion of a narrow region
inside the 2DES \cite{pal,is,man,vk,wharam}; (ii) focused-ion-beam
insulation writing \cite{bev}; (iii) cleaved-edge overgrowth
technique \cite{chang}. As long as the tunnel barrier parameters are
not well controllable values, it is important that using the first
method one can tune the barrier on the same sample. In contrast to
vertical tunneling into the bulk of the 2DES at a quantizing magnetic
field, when the 2DES spectrum shows up \cite{eis,dol}, in the lateral
tunneling electrons can always tunnel into the Landau levels that
bend up at the edge to form edge channels where they intersect the
Fermi level, i.e., the spectrum gaps are not seen directly in lateral
tunneling. Instead, it reflects the edge channel structure and
density of states. For both the integer and fractional quantum Hall
effect, a power-law behaviour of the density of states at the 2DES
edge is expected. This can interfere with the barrier distortion at
an electric field in the nonlinear regime of response and, therefore,
results of lateral tunneling experiments obtained from the
measurements of current-voltage curves \cite{chang} should be treated
with care.

Here, we investigate the lateral tunneling in the narrow
constrictions in which, along a thin strip across, the donor layer of
a GaAs/AlGaAs heterostructure is partly removed using an atomic force
microscope (AFM). A controlled tunnel barrier is created by gate
depletion of the whole of the sample. The well-developed tunneling
regime is indicated by strongly asymmetric diode-like current-voltage
characteristics of the constriction which are sensitive to both gate
voltage $V_g$ and normal magnetic field $B$. The behaviour of the
tunneling part of current-voltage curves points to electron tunneling
between the outermost edge channels.

\begin{figure}[t]
\centerline{
\epsfxsize=\columnwidth
\epsffile{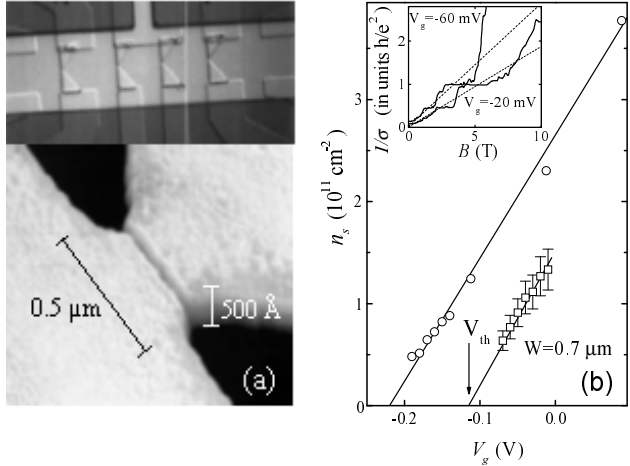}}
\caption{(a) Top view on the sample (top); and a blow-up of one of
the constrictions upon etching the oxidized part of the mesa as
performed solely for visualization purpose (bottom). (b) Gate voltage
dependences of the electron density in the oxidized (squares) and
unoxidized (circles) regions of the 2DES. An example of the
magneto-conductance in the barrier region is shown in the inset. The
value of $n_s$ is extracted from the slope of dashed lines with 10\%
uncertainty.\label{sample}}
\end{figure}

The samples are triangular constrictions of a 2D electron layer with
different widths $W=0.7$, 0.4, 0.3, and 0.2~$\mu$m of the thinnest
part, see Fig.~\ref{sample}(a). These are made using a standard
optical and electron beam lithography from a wafer of GaAs/AlGaAs
heterostructure with low-temperature mobility $\mu=1.6\times
10^6$~cm$^2$/Vs and carrier density $n_s=4\times 10^{11}$~cm$^{-2}$.
Within each constriction the donor layer is removed along a fine line
by locally oxidizing the heterostructure using AFM induced oxidation
\cite{irm}. This technique allows one to define 140~\AA\ wide oxide
lines of sufficient depth and oxide quality so as to partly remove
the donor layer and, therefore, locally decrease the original
electron density. The whole structure is covered with a metallic
gate, which enables us to tune the carrier density everywhere in the
sample. When depleting the 2D layer the oxidized regions get
depopulated first, resulting in the creation of tunnel barriers.
Potential probes are made to the sample to allow transport
measurements.

For the measurements we apply a {\it dc} voltage, $V_{sd}$, between
source (grounded) and drain contacts of one of the constrictions
modulated with small {\it ac} voltage with amplitude
$V_{ac}=40$~$\mu$V and frequency $f=20$~Hz. A gate voltage is applied
between the source and the gate. We measure the real part of the {\it
ac} current, which is proportional to the differential conductance
${\rm d}I/{\rm d}V$, as a function of bias voltage $V_{sd}$ ($I-V$
characteristics) using a home-made $I-V$ converter and a standard
lock-in technique. The behaviour of $I-V$ characteristics is
investigated with changing both gate voltage and magnetic field. The
measurements are performed at a temperature of about 30~mK at
magnetic fields of up to 14~T. The results obtained on different
constrictions are qualitatively similar.

To characterize the sample we extract the gate voltage dependence of
the electron density from the behaviour of magneto-conductance
plateaux in the barrier region and in the rest of the 2DES
(Fig.~\ref{sample}(b)). The analysis is made at high fields where the
size-quantization-caused effect of conductance plateaux in narrow
constrictions is dominated by magnetic field quantization effects
\cite{wees}. As seen from Fig.~\ref{sample}(b), if the barrier region
is depopulated ($V_g< V_{th}$), the electron density in surrounding
areas is still high to provide good conduction. The slopes of the
dependences $n_s(V_g)$ in the oxidized region and in the rest of the
2DES turn out to be equal within our accuracy. The distance between
the gate and the 2DES is determined to be $d\approx 570$~\AA ; as the
corresponding growth parameter is about 400~\AA , the 2D layer
thickness contributes appreciably to the distance $d$. We have found
that even in the unoxidized region the electron density at $V_g=0$
can be different after different coolings of the sample because of
insignificant threshold shifts: it falls within the range $2.5\times
10^{11}$ to $4\times 10^{11}$~cm$^{-2}$ and is always higher compared
to the barrier region.

\begin{figure}[tbh]
\centerline{
\epsfxsize=\columnwidth
\epsffile{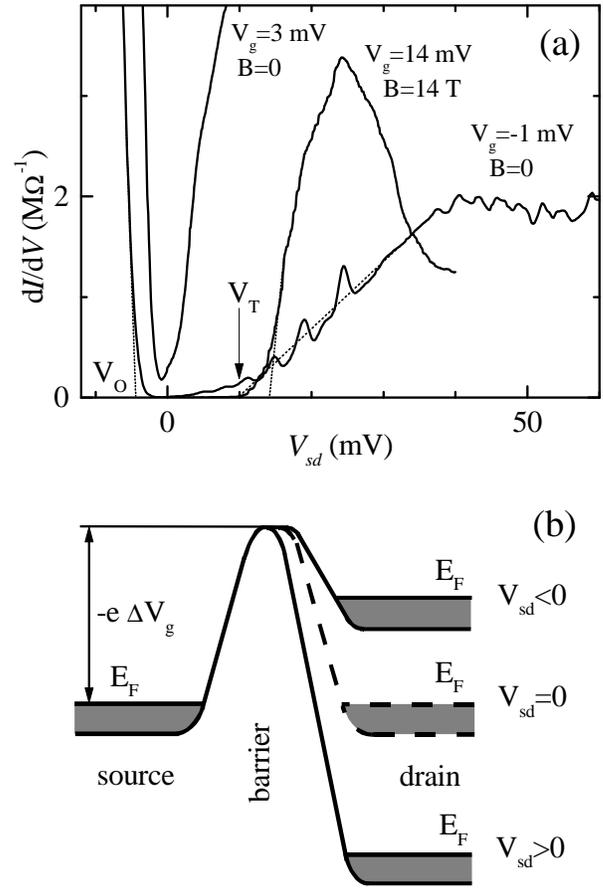}}
\caption{(a) $I-V$ curves at different gate voltages and magnetic
fields. The cases of $B=0$ and $B\neq 0$ correspond to two coolings
of the sample as compared in Fig.~\protect\ref{B0}(a).
$W=0.4$~$\mu$m.
 (b) A sketch
of the 2D band bottom in the barrier region for different
source-drain biases $V_{sd}$.\label{IV}}
\end{figure}

A typical $I-V$ characteristic of the constriction in the
well-developed tunneling regime is strongly asymmetric and includes
an overflowing branch at $V_{sd} <0$ and the tunneling branch at
$V_{sd} >0$, see Fig.~\ref{IV}(a). The tunneling branch is much
smaller and saturates rapidly in zero $B$ with increasing bias
voltage. The onset voltages $V_O$ and $V_T$ for these branches are
defined in a standard way as shown in the figure. The tunneling
regime can be attained by both decreasing the gate voltage and
increasing the magnetic field, as is evident from Fig.~\ref{IV}(a).
We check that the shape of $I-V$ characteristics is not influenced by
interchanging source and drain contacts. Hence, the tunnel barrier is
symmetric and the asymmetry observed is not related to the
constriction geometry.

To understand the asymmetry origin let us consider a gated 2DES
containing a potential barrier of approximately rectangular shape
with width $L\gg d$ in zero magnetic field. The 2D band bottom in the
barrier region coincides with the Fermi level $E_F$ of the 2DES at
$V_g$ equal to the threshold voltage $V_{th}$. Since in the barrier
region for $V_g< V_{th}$ an incremental electric field is not
screened, the 2D band bottom follows the gate potential so that the
barrier height is equal to $-e\Delta V_g=e(V_{th}-V_g)$, where $-e$
is an electron charge (Fig.~\ref{IV}(b)). Applying a bias voltage
$V_{sd}$ leads to shifting the Fermi level in the drain contact by
$-eV_{sd}$. Because of gate screening the voltage $V_{sd}$ drops on
the scale of the order of $d$ near the boundary between barrier and
drain and so the barrier height on the source side does not
practically change, see Fig.~\ref{IV}(b). If $V_{sd}$ reaches the
onset voltage $V_O=\Delta V_g$, the barrier on the drain side
vanishes and electrons start to overflow from drain into source. In
contrast, for $V_{sd} >0$ the electron tunneling through the barrier
from source into drain is only possible. With increasing $V_{sd}$
above $-\Delta V_g$, the tunneling distance diminishes and the
barrier shape becomes close to triangular. Within the triangular
barrier approximation, in the quasiclassical limit of small tunneling
probabilities, it is easy to deduce that the derivative of the
tunneling current over bias voltage is expressed by the relation

\begin{equation}
\frac{{\rm d}I}{{\rm d}V}=\sigma_0\exp\left(-\frac{4(2m)^{1/2}(-e
\Delta V_g)^{3/2}L}{3\hbar eV_{sd}}\right)\ll \sigma_0, \label{eq1}
\end{equation}
where $\sigma_0\approx -(e^2/h)\Delta V_gW/V_{sd}\lambda_F$,
$m=0.067m_0$ ($m_0$ is the free electron mass), and $\lambda_F$ is
the Fermi wave-length in the source. Obviously, the tunneling current
is dominated by electrons in the vicinity of the Fermi level, and the
tunneling distance $L_T=-\Delta V_gL/V_{sd}$ should satisfy the
inequality $d\ll L_T <L$. In accordance with Eq.~(\ref{eq1}), the
expected dependence of the tunneling onset voltage $V_T$ on gate
voltage is given by $V_T\propto (-\Delta V_g)^{3/2}$.

\begin{figure}[t]
\centerline{
\epsfxsize=\columnwidth
\epsffile{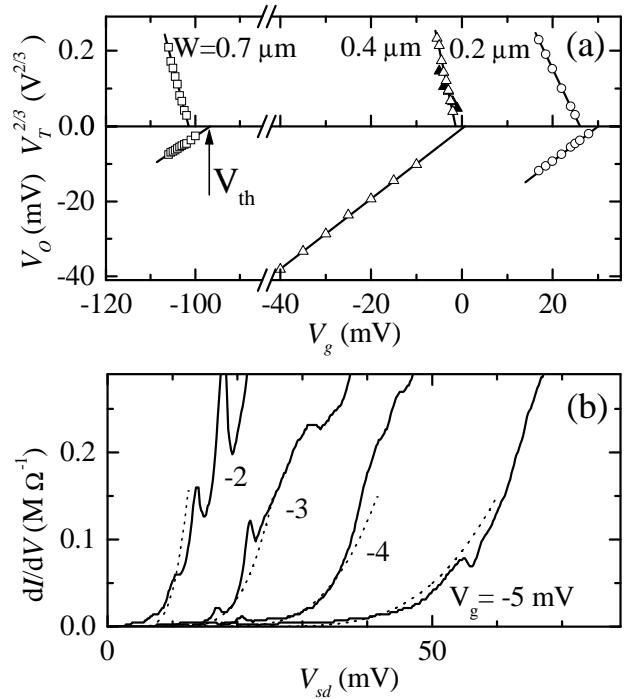}}
\caption{(a) Change of the onset voltages $V_O$ and $V_T$ as defined
in Fig.~\protect\ref{IV}(a) with $V_g$ at $B=0$; and (b) the fit
(dashed lines) of $I-V$ curves (solid lines) by
Eq.~(\protect\ref{eq1}) with the parameters $L=0.6$~$\mu$m,
$\sigma_0=38$~$M \Omega^{-1}$,
$V_{th}=-0.4$~mV;
$W=0.4$~$\mu$m.
 In case (a),
the data marked by filled triangles are obtained for the same cooling
of the sample as the ones at $B=0$ in Fig.~\protect\ref{IV}(a) and in
case (b), whereas the open triangles correspond to the $B\neq 0$ data
of Figs.~\protect\ref{IV}(a) and \protect\ref{strongB} as measured
for the other cooling.\label{B0}}
\end{figure}

As seen from Fig.~\ref{B0}(a), the expected behaviour of both $V_O$
and $V_T$ with changing $V_g$ is indeed the case. The dependences
$V_O(V_g)$ and $V_T^{2/3}(V_g)$ are both linear; the slope of the
former is very close to one. Extensions of these straight lines
intercept the $V_g$-axis at slightly different voltages, which points
out that the triangular barrier approximation is good. The threshold
voltage $V_{th}$ for the 2DES's generation in the barrier region,
which is defined as a point of vanishing $V_O$ (Fig.~\ref{B0}(a)), is
coincident, within experimental uncertainty, with the value of
$V_{th}$ determined from the analysis of magneto-conductance plateaux
(Fig.~\ref{sample}(b)).

Fitting the set of $I-V$ characteristics at different $V_g$ by
Eq.~(\ref{eq1}) with parameters $L$, $V_{th}$, and $\sigma_0$ is
depicted in Fig.~\ref{B0}(b). The dependence of $\sigma_0$ on $\Delta
V_g$ and $V_{sd}$ is ignored on the background of the strong
exponential dependence of ${\rm d}I/{\rm d}V$. Although three
parameters are varied, the fit is very sensitive, except for
$\sigma_0$, to their variation because of the exponential behaviour
of $I-V$ characteristics. One can see from Fig.~\ref{B0}(b) that the
above model describes well the experiment at zero magnetic field. As
expected, the determined parameter $L=0.6$~$\mu$m is much larger than
$d$, i.e. the barrier shape at $V_{sd}=0$ is approximately
rectangular, and the value of $V_{th}$ is close to the point where
$V_O$ (and $V_T$) tends to zero (Fig.~\ref{B0}(a)). The similar
results are obtained on the other two constrictions. Besides, we find
that the coefficient $\sigma_0$ for different constrictions does not
scale with the constriction width $W$. This probably implies that the
tunnel barriers even with submicron lengths are still inhomogeneous,
which, however, does not seem crucial for the case of exponential
$I-V$ dependences.

\begin{figure}[tb]
\centerline{
\epsfxsize=\columnwidth
\epsffile{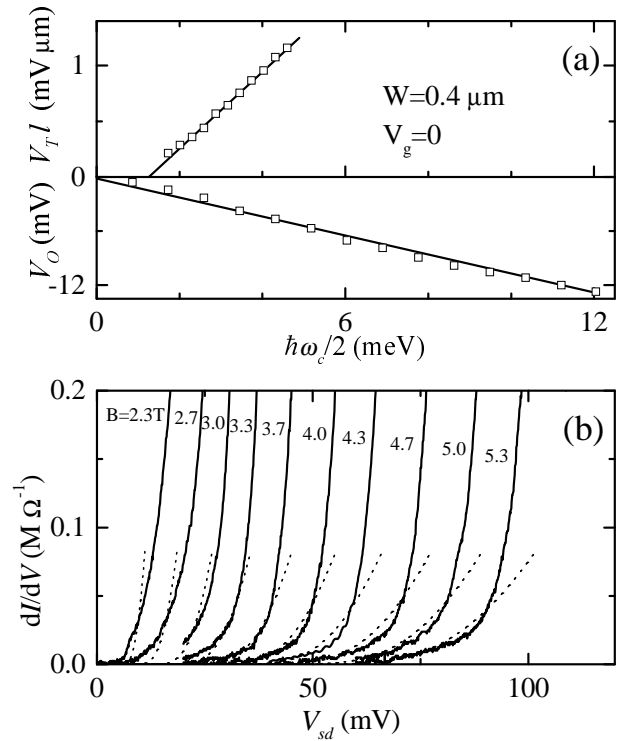} }
\caption{(a) Behaviour of the onset voltages $V_O$ and $V_T$ with
magnetic field; and (b) the fit (dashed lines) of $I-V$ curves (solid
lines) by Eq.~(\protect\ref{eq4}) with the parameters
$L=0.6$~$\mu$m,
$\sigma_0=1.3$~$M \Omega^{-1}$,
$V_{th}=-1.4$~mV;
$W=0.4$~$\mu$m,
$V_g=0$.
\label{strongB}}
\end{figure}

Having tested that we deal with the controlled tunnel barrier we
investigate the tunneling at a normal magnetic field that gives rise
to emerging the tunnel barrier in a similar way to gate depletion
(Fig.~\ref{IV}(a)). At constant $V_g> V_{th}$, where there is no
tunnel barrier in zero $B$, the magneto-conductance $\sigma$ obeys
the $1/B$ law at weak fields and drops exponentially with $B$ in the
high-field limit, signaling the tunneling regime.
Fig.~\ref{strongB}(a) presents the magnetic field dependence of the
onset voltage $V_O$ that determines the barrier height. As seen from
the figure, the change of the barrier height $-eV_O$ with $B$ is very
close to $\hbar\omega_c/2$, which points to a shift of the 2D band
bottom by half of the cyclotron energy.

For describing the tunneling branch of $I-V$ characteristics we
calculate the tunneling probability in the presence of a magnetic
field. This is not so trivial as at $B=0$ because electrons tunnel
through the magnetic parabola between edge channels at the induced
edges of the 2DES. In the triangular barrier approximation one has to
solve the Schr\"odinger equation with the barrier potential

\begin{equation}
U(x)=\frac{\hbar\omega_c}{2l^2}(x-x_0)^2-eV_{sd}\frac{x}{L}-e\Delta
V_g, \mbox{ } 0<x<L, \label{eq2}\end{equation}
where $\omega_c$ is the cyclotron frequency, $l$ is the magnetic
length, and $eV_{sd}$ is larger than the barrier height in the
magnetic field. An electron at the Fermi level in the source tunnels
through $U(x)$ from the origin to a state with orbit centre $x_0$
such that $0<x_0<L$. If the barrier potential is dominated by the
magnetic parabola (i.e., the magnetic length is the shortest), the
problem reduces to the known one to find energy levels in the shifted
parabolic potential as caused by the linear term in Eq.~(\ref{eq2}).
The value of $x_0$ is determined from the condition of coincidence of
a Landau level in the potential $U(x)$ with the Fermi level in the
source. If the lowest Landau level is regarded only and the spin
splitting is ignored, we get the minimum tunneling distance to the
outermost edge channel in the drain

\begin{equation}
x_0=L_T=\frac{l}{2}\left(\frac{\hbar\omega_cL}{eV_{sd}l}-
\frac{2\Delta V_gL}{V_{sd}l}-\frac{eV_{sd}l}{\hbar\omega_cL}\right)
\gg d. \label{eq3}\end{equation}
The first term in brackets in Eq.~(\ref{eq3}), which is dominant, is
large compared to one. Knowing the wave function of the lowest Landau
level in the potential $U(x)$ and neglecting the last term in
Eq.~(\ref{eq3}) we obtain for the shape of $I-V$ characteristics near
the onset, where the tunneling probability is small,

\begin{equation}
\frac{{\rm d}I}{{\rm d}V}=\sigma_B\exp\left(-\frac{(\hbar\omega_c/2
-e\Delta V_g)^2L^2}{e^2V_{sd}^2l^2}\right)\ll\sigma_B. \label{eq4}
\end{equation}
Here $\sigma_B$ is the pre-factor which can be tentatively expected
to be of the same order of magnitude as $\sigma_0$. From
Eq.~(\ref{eq4}) it follows that at sufficiently strong magnetic
fields the tunneling onset voltage $V_T$ is related to the barrier
height as $V_Tl\propto \hbar\omega_c/2- e\Delta V_g$, which is
consistent with the experiment (Fig.~\ref{strongB}(a)). The solution
(\ref{eq4}) includes the case $e\Delta V_g >0$ when a tunnel barrier
is absent at zero magnetic field but arises with increasing $B$. This
occurs apparently because of depopulation of the barrier region in
the extreme quantum limit of magnetic field.

Fig.~\ref{strongB}(b) displays the fit of $I-V$ characteristics at
different magnetic fields by Eq.~(\ref{eq4}) with parameters $L$,
$V_{th}$, and $\sigma_B$. The optimum values of $L=0.6$~$\mu$m and
$V_{th}=-1.5$~mV are found to be very close to the ones for the $B=0$
case as determined for the same range of barrier heights, see
Fig.~\ref{B0}(b). Although this fact supports our considerations,
these are not quite rigorous to discuss the considerable discrepancy
between the fore-exponential factors with and without magnetic field.

The observed behaviour of $I-V$ characteristics with magnetic field
in the transient region where their asymmetry is not yet strong
(Fig.~\ref{IV}(a)) is similar to that of Refs.~\cite{vk,chang}. Over
this region, which is next to the scope of the exponential $I-V$
dependences at higher magnetic-field-induced tunnel barriers, our
$I-V$ curves are close to power-law dependences, as was discussed in
Ref.~\cite{chang}. There is little doubt that it is very difficult to
analyze and interpret such $I-V$ curves without solving the tunneling
problem strictly. We note that the peak structures on the tunneling
branch of $I-V$ characteristics (see Figs.~\ref{IV}(a) and
\ref{B0}(b)) persist at relatively low magnetic fields and are very
similar to those studied in Ref.~\cite{vk}. These may be a hint to
resonant tunneling through impurity states below the 2D band bottom.

This work was supported in part by the Russian Foundation for Basic
Research under Grants No.~97-02-16829 and No.~98-02-16632, the
Programme "Nanostructures" from the Russian Ministry of Sciences
under Grant No.~97-1024, and Volkswagen-Stiftung under Grant
No.~I/68769.

\end{document}